\documentclass[a4paper, 12pt]{article}
\usepackage{graphicx}
\usepackage{amsmath}
\usepackage{amssymb}

\def \be {\begin{equation}}
\def \ee {\end{equation}}
\def \bea {\begin{eqnarray}}  
\def \eea {\end{eqnarray}}  
\def \mea {\nonumber\\}
\def \lammax {{\lambda_{\rm {max.}}}}

\def \Qop {\widehat{\Delta}_{(0,\,T)}}

\usepackage{cite}

\begin{document}
\title{Remarks on Quantum Probability Backflow}     
     
\author{ A.J. Bracken\footnote{{\em Email:} a.bracken@uq.edu.au}  \\School of Mathematics and Physics\\    
University of Queensland\\ Brisbane 4072, Australia\\ and\\ J.B. McGuire
\footnote{{\em Email:} mcguirej@fau.edu}
\\
Department of Physics\\ Florida Atlantic University\\ Boca Raton, FL 33431--0991, USA  }
\date{} 
\maketitle     

\begin{abstract}

It is known that for a non-relativistic quantum particle traveling freely on the $x$-axis,
the positional probability can flow in the opposite direction to the particle's velocity. 
The maximum possible amount of such backflow that can occur over any time interval has been determined previously as the 
largest positive eigenvalue of a certain  hermitian observable, with the value  $0.0384517\dots$, or about $4\%$ of the total probability 
on the line.  
The eigenvalue problem is now considered numerically in the more general case of states with momentum restricted to the range
$p_0<p<\infty$, for any given value $p_0$.  It is found that the maximum possible backflow decreases monotonically, but
never reaches $0$, as $p_0$ 
increases through positive values; and it increases monotonically, but never reaches $1$, as $p_0$ decreases through negative values.  Both of these effects 
are non-classical.  The results allow a simple interpretation of the classical limit, as an effective value of Planck's constant goes to zero
and probability backflow becomes impossible.  
\end{abstract}


\section{Introduction}
A  consequence of wave-particle duality is that probability backflow can occur during the motion 
of a quantum `particle', in particular one moving freely on a straight line, say the $x$-axis, and 
described by a wave packet composed
of plane waves, each of which has positive wave number (momentum)  \cite{allcock,bracken1,bracken4,nielsen}.  
Despite the fact that a particle in such a state would {\em certainly}  
be found on measurement of its  momentum to be moving with constant velocity from left to right, the 
probability of finding it to the left of a given point on the axis, say $x=0$, on measurement of its  position, can
increase over any given time interval.  Thus it seems that the positional probability can flow in the opposite
direction to the velocity of the particle.  

It is important to see at the outset that the backflow phenomenon is not simply the familiar ``spreading
of a quantum wave packet'',  such as a Gaussian packet with a constant {\em average} velocity,  directed 
from  left to right.  Such a packet  spreads in both directions as it travels, but that is not in itself surprising, 
because the packet is composed of left-moving as well as right-moving plane waves, so that there is a finite 
probability that the particle will be found on measurement of its momentum to be traveling in either direction.  
The backflow surprise is greatest when the probability flows in one direction even though the 
particle is --- more precisely, would be found on measurement to be --- certainly traveling in the opposite direction.

The reader may well consider that peculiar quantum 
effects involving incompatible observables 
such as position and momentum
are to be expected.  Furthermore, a quantum object has wave-like as well as particle-like properties and, from 
a purely mathematical point of view, probability backflow is one more example of retrogressive wave motion 
and is not particularly remarkable as such \cite{berry}.  
 Nevertheless,  in the context of  quantum 
 dynamics   the effect is strikingly counterintuitive, with no classical analogue.  
 
By what `surreal  behavior'  can a `particle' with velocity
directed left-to-right, increase its chance of being found to the left of a given point?  
To make a connection with interpretations of other
`surreal' quantum  effects in terms of Bohmian mechanics  \cite{englert,durr,becker,scully,hiley}, 
it is enough here  to note that while probability backflow is occurring,
the Bohmian velocity \cite{bohm,wiseman}, by definition 
parallel to the probability flux vector, is directed opposite to any velocity value 
obtained by measuring the 
momentum observable.  

A quite different interpretation \cite{bracken1,bracken4,nielsen} follows Feynman's  suggested  
introduction of negative probabilities \cite{dirac,feynman} as an aid to 
the  interpretation of aspects
of quantum behavior that defy a classical explanation,  such as those associated with the double-slit experiment.  
Negative probabilities appear 
through the introduction of 
the Wigner function \cite{wigner}, leading in the present context to the notion that while all probability 
flows in the same direction as the velocity 
for a quantum particle, just as  for a
classical particle,  in the quantum case the flowing probability can be negative during some time intervals.  The flow of
negative probability in the direction of the velocity produces the same effect as a flow of positive 
probability in the opposite direction --- probability backflow.

The possibility of probability backflow was first noticed over fifty years ago 
in the case of a free, non-relativistic particle  \cite{allcock}, and the phenomenon was subsequently described in detail and 
quantified for that case \cite{bracken1}.  It has been widely discussed and generalized in various directions since \cite{bracken4,nielsen,eveson,penz,yearsley,halliwell,strange,bracken2,bracken3,muga3,leavens}.  
The maximum size   of the effect is small \cite{bracken1,penz,eveson}; in  the case of a free particle,  
no more than $0.0384517\dots$ out of the total (unit) probability on the line, 
or about $4\%$, can flow backwards in any given time interval of length $T$, no matter how short or long. 

The properties of this `quantum number' $\lammax\approx 0.04$ provide another reason for interest in probability backflow.  
It is
not only dimensionless and independent of the length of the time interval involved, 
but is also independent of the size of Planck's constant $\hbar$
and of the mass $m$ of the particle \cite{bracken1}.
Experimental measurement of its value might provide an unusual new test of the structure of quantum dynamics.

A possible experiment to observe probability backflow  
has been devised recently \cite{muga1,muga2}, but to our knowledge no experiment has yet been 
proposed to go further and  confirm the predicted maximum size $\lammax$ of the effect. 
\section{Purpose}

 The purpose of this note is
 to throw more light
 on the phenomenon of probability backflow  in the case of a free, non-relativistic particle, by considering the mathematical question:  
 How does the maximum backflow value $\lammax$ change
 as the least value of the momenta in the waves making up a packet  is varied?  In other words, what is the 
 maximum backflow that can occur over a time interval of length $T$  
 if packets containing only momenta $p>p_0$ are considered, 
 for given $-\infty < p_0<\infty$?  

 When the choice $p_0=0$ is made, then $\lammax = 0.0384517\dots$
 as indicated above.   It
 is to be expected that $\lammax$ will decrease towards $0$ as $p_0$ increases, because probability backflow
 surely becomes more and more unlikely as the minimum left-to-right velocity of the particle is increased.  Conversely,
  as $p_0$ decreases below $0$, the flow of some probability to the left is no longer counterintuitive,
  and it is to be expected that $\lammax$ will increase towards $1$, representing the total
  probability on the line.  But the shape of the graph of $\lammax$ {\em v.}  $p_0$
 for $-\infty<p_0<\infty$ is quite unclear.   At present the only known point on the graph is 
$(p_0=0,\lammax =0.0384517\dots)$.   The object here is to present and discuss a numerically-determined approximation to this graph.

\section{The graph in question}
For a  non-relativistic particle with mass $m$ moving freely on the $x$-axis, 
the value of $\lammax$ is determined from the eigenvalue problem for the hermitian integral operator
$\Qop $ that acts on momentum-space amplitudes $\phi(p)$ as \cite{bracken1}
\bea
(\Qop\, \phi )(p)=\frac{i}{2\pi}\,\int\,
\frac{e^{i (p^2-q^2)\,T/2m\hbar}-1}{p-q}\,\phi(q)\,dq\,.
\label{Qaction}
\eea

This `probability-flow operator' is constructed \cite{bracken1} from the probability flux vector, and 
as a quantum observable has the meaning that its expectation value in a given quantum state 
of the  particle at time $t$, is the amount of probability that will flow across $x=0$ from right to left in the time-interval
$[t,\,t+T]$ during the free-particle evolution.  Accordingly, the largest possible probability backflow
during that time interval
is given by  the largest eigenvalue of $\Qop$.  More precisely,
since it is not certain that the spectrum of $\Qop$ is discrete \cite{penz}, $\lammax$  is the least upper bound
on that spectrum.  
Note that $\Qop$  is independent of $t$ and hence the same is true of $\lammax$.

The range of the variable $p$ and of the integration over $q$ in \eqref{Qaction},  both equal some chosen range of 
variation of momentum values.  In the problem studied so far, as outlined in the Introduction, 
only amplitudes $\phi(p)$ are considered where the range of integration and of $p$ is
$0<p<\infty$.  In states described by such amplitudes, 
the particle is certain to be found on measurement to have a positive momentum.
 In what follows, the more general range $p_0<p<\infty$ is considered, for various choices of $-\infty<p_0<\infty$.  
 For each such choice, the momentum of the particle 
would certainly be found on measurement
to lie in this more general range, between $p_0$ and $\infty$.  

The more general eigenvalue problem can be written in dimensionless form as
\bea
-\frac{1}{\pi}\,\int_{u_0}^\infty \, \frac{\sin (u^2-v^2)}{u-v}\,\varphi(v)\,dv =\lambda\,\varphi(u)\,,\quad 
u_0<u<\infty\,,
\label{evals1}
\eea
where
\bea
\varphi(u)=e^{-ip^2T/4m\hbar}\,\phi(p) \,,\quad \varphi(v)=e^{-iq^2T/4m\hbar}\,\phi(q) \,,
\mea\mea
u=\sqrt{\frac{T}{4m\hbar}}p\,,\quad v=\sqrt{\frac{T}{4m\hbar}}q\,,\quad u_0=\sqrt{\frac{T}{4m\hbar}}p_0\,.
\label{evals2}
\eea
It is clear from \eqref{evals1} that the spectrum of $\lambda$ values, and so the value of $\lammax$  in this general  case, is
dependent on the values of $p_0$,  $T$, $\hbar$ and $m$ only through the dimensionless parameter
$u_0$ as in \eqref{evals2}.  When $p_0=0$, so that $u_0=0$,  it is independent of $T$, $\hbar$ and $m$ as already noted above.

Fig. 1 shows the interesting part of the  graph of $\lammax$ {\em v.} $u_0$, obtained numerically by discretizing a sufficiently
large part of the range of $u$ and $v$ values in \eqref{evals1} and using the MATLAB \cite{matlab}
eigenvalue solver {\em eigs} repeatedly.

\begin{figure}[ht]
\centering
\includegraphics[width=7in,angle=90]
{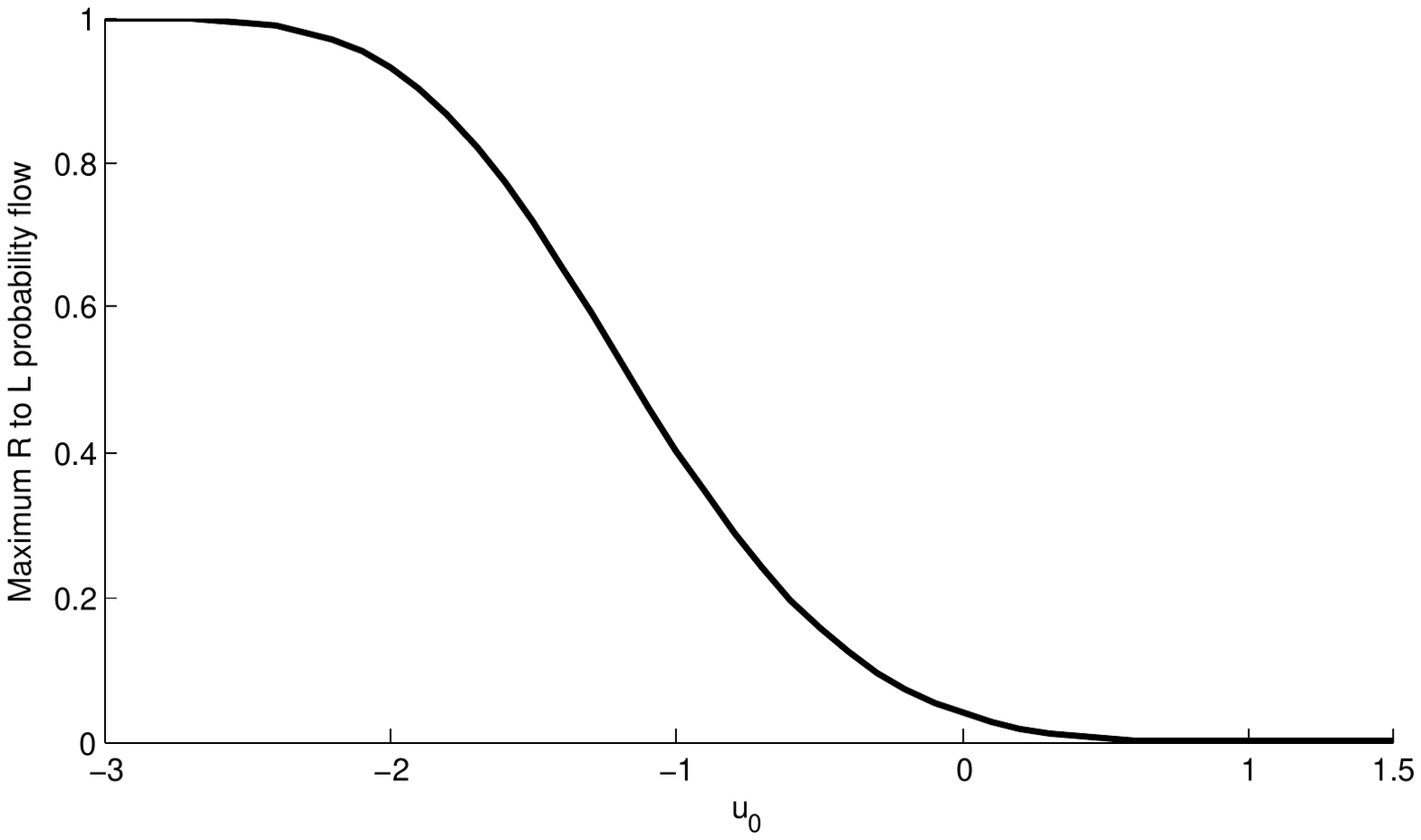}
\caption{Maximum R to L probability flow $\lammax$  plotted against $u_0$. Note $\lammax\approx 0.04$ when $u_0=0$. }
\end{figure}

There are two surprises.  The first is that no matter how large {\em positive} is $u_0$ as in \eqref{evals2}, so no matter how large
positive is
the minimum cut-off  momentum value $p_0$ for given $T$, $\hbar$ and $m$,  some probability backflow can still occur,
though $\lammax (u_0)$, the maximum amount of backflow possible for given $u_0$, 
decreases steadily towards $0$ as $u_0$ increases.  This result,
and the way in which $\lammax (u_0)$ depends on the values of $T$ and $m$ through $u_0$ as in \eqref{evals2},  may be
useful in the design of  experiments like that mentioned in the Introduction \cite{muga1,muga2}.

The second surprise is that no matter how large {\em negative} is $u_0$, so no matter how large
negative is
the minimum cut-off  momentum value $p_0$ for given $T$, $\hbar$ and $m$,  the maximum amount $\lammax$ of right-to-left 
probability flow  is less than 1, though it increases steadily towards $1$ as $u_0$ decreases. 

Both these results are non-classical.   For a classical particle 
traveling along the $x$-axis with  an uncertain position and
 a constant but uncertain momentum $p>p_0$, the corresponding graph is discontinuous, and consists of two
straight  line segments, namely $\lammax=1$ for $p_0<0$, and  $\lammax=0$ for $p_0>0$.  
The result for $p_0>0$ is obviously true; probability backflow cannot occur for a classical particle.  
To see that the result for $p_0<0$  is also correct, it suffices to
consider the evolution of a classical probability density for momentum values that 
has compact support bounded below by $p_0$ and 
contained entirely on the negative $p$-axis, with in addition  an initial probability density for position 
values with compact support entirely 
on the positive  $x$-axis. Then the right-to-left probability flow
across $x=0$
in an appropriate time interval is clearly $1$, the  maximum possible.   

Discussions of the `classical limit' of probability backflow  \cite{bracken1,yearsley} when $p_0=0$ have faced the obstacle
that $\lammax$ is independent of $\hbar$ in that case, as noted in the Introduction.  But the broadening of the treatment of the eigenvalue problem 
for $\Qop$ by considering each $-\infty<p_0<\infty$ leads to an appealing resolution of this difficulty. 
The discontinuous classical curve can be obtained in the  `classical limit' from the quantum curve
by considering the graph of $\lammax (u_0/\alpha)$ {\em v.} $u_0$, 
effectively replacing
$\hbar$ by $\hbar_{\rm {eff.}}=\alpha^2\hbar$ in \eqref{evals2}, and then allowing $\alpha\to 0$.  Fig. 2 shows the
graphs for $\alpha=1$, $3/4$, $1/2$, $1/4$ and $1/8$.  Note that all the curves pass through the same value
$\lammax\approx 0.04$ at $u_0=0$;  the maximum probability backflow in that previously-studied 
case is indeed independent of the value of $\hbar$, and does not change as $\hbar_{\rm {eff.}}\to 0$, while the quantum curves
nevertheless approach the discontinuous classical curve.

\begin{figure}[ht]
\centering
\includegraphics[width=7in,angle=90]
{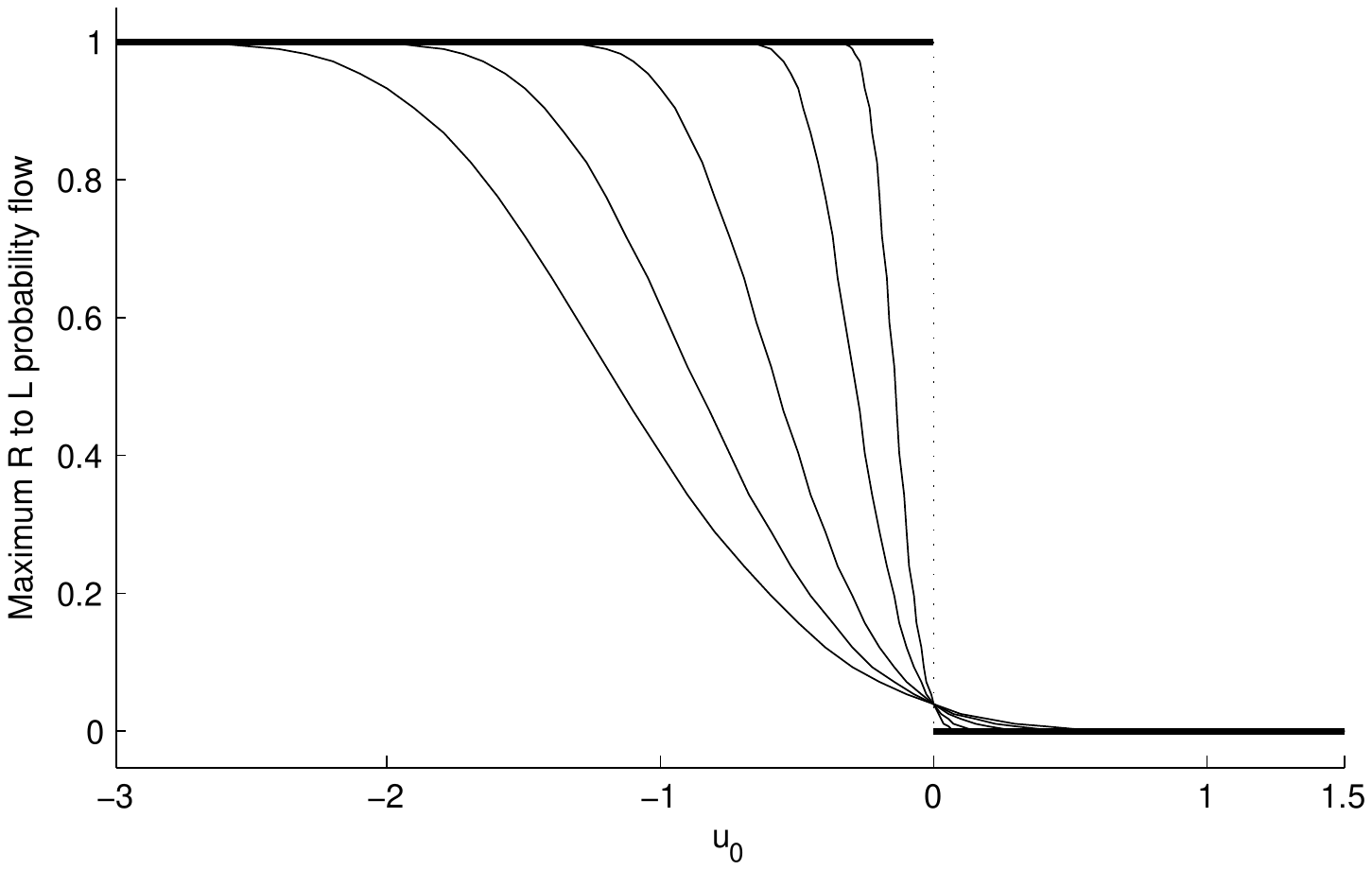}
\caption{Maximum R to L probability flow plotted against $u_0$ for $\hbar_{\rm {eff.}}=\alpha^2 \hbar$,
with $\alpha =1,\,3/4,\,1/2,\,1/4,\,1/8$ from left to right at the top, 
showing approach to the discontinuous `classical limit' curve as $\hbar_{\rm {eff.}}\to 0$. }
\end{figure}

A final remark: A cursory inspection  of the curve in Fig. 1 suggests that the function $\lammax(u_0)-0.5$ may be 
odd about the point $u_0=u_0^*\approx -1.16$ where $\lammax-0.5=0$, that is to say,  that
\bea
\lammax(u_0-u_0^*)-0.5=- (\lammax(u_0^*-u_0)-0.5)\,,
\label{evals3}
\eea
but a more accurate numerical study is needed before this can be claimed true with confidence.
Such a result would imply an important symmetry of the eigenvalue problem \eqref{evals1}.

\end{document}